2

\documentclass[twocolumn]{aastex61}

\accepted{April 9, 2018}
\submitjournal{ApJL}

\shorttitle{The influence of stellar spin on ignition of thermonuclear runaways}
\shortauthors{Galloway et al.}

\begin{document}

\title{The influence of stellar spin on ignition of thermonuclear runaways}

\correspondingauthor{Duncan Galloway}
\email{duncan.galloway@monash.edu}

\author[0000-0002-6558-5121]{Duncan K. Galloway}
\affil{School of Physics \& Astronomy, 
Monash University, 
Clayton, VIC 3800, Australia}
\affiliation{Monash Centre for Astrophysics, 
Monash University, 
Clayton, VIC 3800, Australia}

\author{Jean J. M. in 't Zand}
\affiliation{SRON Netherlands Institute for Space Research, 
Sorbonnelaan 2, 3584 CA Utrecht, the Netherlands}

\author{J\'er\^ome Chenevez}
\affiliation{National Space Institute, 
 Technical University of Denmark, 
  Elektrovej 327-328, DK-2800 Lyngby, Denmark }

\author{Laurens Keek}
\affiliation{X-ray Astrophysics Laboratory, 
 Astrophysics Science Division, 
  NASA/GSFC, Greenbelt, MD 20771}
\affiliation{CRESST and the Department of Astronomy, 
 University of Maryland,
  College Park, MD 20742}

\author{Celia Sanchez-Fernandez}
\affiliation{European Space Astronomy Centre (ESA/ESAC), 
 Science Operations Department, 
  E-28691, Villanueva de la Ca\~nada, Madrid, Spain}

\author{Hauke Worpel}
\affiliation{Leibniz-Institut f\"ur Astrophysik Potsdam (AIP), 
 An der Sternwarte 16, Potsdam 14482, Germany}
 
\author{Nathanael Lampe}
\affiliation{CNRS, IN2P3, CENBG, UMR 5797, F-33170 Gradignan, France}

\author{Erik Kuulkers}
\affiliation{ESA/ESTEC, Keplerlaan 1, 2201 AZ Noordwijk, The
Netherlands}

\author{Anna Watts}
\affiliation{Anton Pannekoek Institute for Astronomy, University of
Amsterdam, Postbus 94249, 1090GE Amsterdam, the Netherlands},

\author{Laura Ootes}
\affiliation{Anton Pannekoek Institute for Astronomy, University of
Amsterdam, Postbus 94249, 1090GE Amsterdam, the Netherlands},
\collaboration{The MINBAR collaboration}

\begin{abstract}

Runaway thermonuclear burning of a layer of accumulated fuel on the surface of a compact star provides a brief but intense display of stellar nuclear processes. 
For neutron stars accreting from a binary companion, these events manifest as thermonuclear (type-I) X-ray bursts, and recur on typical timescales of hours to days.
We measured the burst rate as a function of accretion rate, from seven neutron stars with known spin rates, using a burst sample accumulated over several decades. At the highest accretion rates, the burst rate is lower for faster spinning stars. The observations imply that fast ($>400$~Hz) rotation encourages stabilization of nuclear burning, suggesting a dynamical dependence of nuclear ignition on the spin rate. This dependence is unexpected, because faster rotation entails less shear between the surrounding accretion disk and the star. 
Large-scale circulation in the fuel layer, leading to enhanced mixing of the burst ashes into the fuel layer, may explain this behavior;  further numerical simulations are required to confirm.

\end{abstract}


\section{Introduction} \label{sec:intro}

Thermonuclear runaways arise from unstable ignition of accreted fuel on
the surface of compact objects in binary systems. One such class of events
has been known since antiquity as novae \cite[]{jose07}, and occur on the
surface of white dwarfs accreting from a stellar companion. A second class
of runaway was discovered in the early 1970s with the advent of
satellite-based X-ray astronomy \cite[]{grindlay76,clark76}, and occurs on
neutron stars, where the surface temperature is sufficiently hot that the
star emits primarily in X-rays. These X-ray bursts occur typically $10^6$
times more frequently than novae, with recurrence times of hours rather
than years. At typical accretion rates ($\sim0.1\ \dot{M}_{\rm Edd}$,
where $\dot{M}_{\rm Edd}=1.3\times10^{-8}\ M_\odot\,{\rm yr^{-1}}$ is the Eddington limit for a $1.4\ M_\odot$ object accreting solar composition fuel), bursts occur quasi-periodically every few hours, with durations of 10--100~s \cite[e.g.][]{lew93,gal17b}. 

Few recurrent novae are known \cite[e.g.][]{darnley16}, although the
population has expanded dramatically in the last decade, thanks primarily
to surveys of other galaxies. The relative frequency of runaways on
neutron stars enables much more detailed studies than for novae. Most of
the $\approx100$ known X-ray burst
sources\footnote{\url{http://burst.sci.monash.edu/sources}} within our
Galaxy have shown multiple bursts, with over a thousand events detected
from the most prolific sources. In addition, many of the bursting neutron
stars have been shown to be spinning with spin frequencies in the range
11--620~Hz \cite[]{watts12a}, introducing the possibility of studying rotational effects that may influence the spreading of the thermonuclear flame.

The influence of stellar spin on thermonuclear ignition has been investigated theoretically via its effect on mixing \cite[]{fuji93}. At the depth of thermonuclear burning on neutron stars (column depth $y\sim10^8\ {\rm g\,cm^{-2}}$) viscosity likely results in nearly uniform rotation. Even so, turbulent mixing can occur, particularly at high accretion rates \cite[]{pb07}. Numerical simulations suggest that this mixing may stabilize the burning at accretion rates lower than would be expected in nonrotating layers \cite[]{keek09}. However, most calculations have focussed on pure helium layers, while many neutron stars accrete both hydrogen and helium.

Another factor that may influence the conditions in the fuel layer is the shear with respect to the accretion flow. If the accretion disk meets the neutron star surface forming a boundary layer \cite[as is thought to be the case at high accretion rates;][]{done07} then the shear between the surface and the disk material (orbiting at roughly Keplerian speeds) may contribute heating even into the fuel layer \cite[]{is10}. Because the Keplerian motion of the disk is always faster than the rotation speed of the neutron star, the shear heating can be expected to have a greater effect for slower neutron star spin; in contrast, rotation-induced mixing will have a greater effect at {\it faster} spin rates.

Observationally, variations in the persistent spectral shape as a function of spin rate have been attributed to the effects of shear heating  \cite[]{burke18}.
The spin rate also seems to determine in part the type of bursts in which ``burst oscillations'' \cite[]{watts12a} are observed. For rapidly spinning sources ($\nu_{\rm spin}\gtrsim500$~Hz) burst oscillations are predominantly observed in bursts exhibiting so-called photospheric radius-expansion \cite[PRE;][]{muno04a}. The X-ray intensity during such events reaches the local Eddington limit, where the force due to radiation pressure from the burst emission balances the (considerable) force due to gravity at the neutron star surface. In contrast, burst oscillations in slower-rotating sources are prevalent equally likely in bursts with or without PRE. This effect may arise partially from the predominance in the slower-spinning sources for (almost) pure He accretion \cite[]{bcatalog,gal10b}.

Here we present a study of the variation in the rate of thermonuclear
runaways in response to the accretion rate in different sources, to determine the influence of the spin rate on thermonuclear ignition.

\section{Observations and analysis} 
\label{sec:obs}

We employ the largest
available sample of thermonuclear bursts, incorporating more than 7000 
events from 85 bursting sources, observed by three long-duration satellite
X-ray missions: the Dutch-Italian {\it BeppoSAX},
NASA's {\it Rossi X-ray
Timing Explorer}\/ ({\it RXTE}), and ESA's {\it INTEGRAL}\/ satellites.
The sample, comprising the Multi-INstrument Burst ARchive
(MINBAR\footnote{\url{http://burst.sci.monash.edu/minbar}}), includes uniform analyses of the burst properties, as well as the persistent (non-bursting) X-ray emission, from which the rate of accretion onto the neutron star can be estimated.

We carried out our analysis for seven sources for which there are $\approx100$ or more bursts in MINBAR, and for which the spin frequency is known (Table \ref{tab:sources}).
The spin frequency for these sources has been inferred from the detection of burst oscillations \cite[e.g.][]{watts12a}, which are known to trace the neutron star spin \cite[]{chak03a}.

\begin{deluxetable*}{lcccCCc}
\tablecaption{Bursting sources selected for burst rate measurements
\label{tab:sources}}
\tablecolumns{4}
\tablewidth{0pt}
\tablehead{
& \colhead{Spin frequency} &
\colhead{No.} & \colhead{Exposure} & 
\colhead{$\left<F_{\rm Edd}\right>$} & &
 \\
\colhead{Source} & \colhead{(Hz)} & \colhead{bursts} & \colhead{(Ms)} & 
\colhead{($10^{-8}\ {\rm erg\,cm^{-2}\,s^{-1}}$)} & \colhead{$c_{\rm bol}$}
& \colhead{$\xi_p$ ($\xi_b$)} }
\startdata
4U 1702$-$429 &  329 &  280 &   7.22 & 8.8\pm0.5 & 1.62\pm0.13 & 0.809 (0.898)\\
4U 1728$-$34 &  363 & 1172 &  12.47 & 9.5\pm0.8 & 1.47\pm0.05 & 0.809 (0.898)\\
KS 1731$-$260 &  524 &  369 &   5.01 & 4.9\pm0.6 & 1.77\pm0.09 & 0.809 (0.898)\\
Aql X-1 &  550 &   96 &   2.64 & 10\pm2 & 2.10\pm0.17 & 0.809 (0.898)\\
EXO 0748$-$676 &  552 &  357 &   5.98 & 4.7\pm0.5 & 1.563\pm0.009 & 3.21 (1.40)\\
4U 1636$-$536 &  581 &  666 &   8.57 &  7.2\pm0.9 & 1.80\pm0.05 & 0.809 (0.898)\\
4U 1608$-$522 &  620 &  146 &   5.08 & 17\pm3 & 1.58\pm0.09 & 0.809 (0.898)\\
\enddata
\end{deluxetable*}

We measured the (average) burst rate as a function of accretion rate, as
follows.
We estimated the accretion rate for each observation from the intensity of the non-bursting (persistent) emission, averaged over each observation, largely following \cite{bcatalog}. We then grouped the bursts detected within each observation to measure the burst rate at different accretion rates. 

The instantaneous accretion rate onto bursting neutron stars can be inferred from measurements of the persistent (non-bursting) X-ray emission, and varies on timescales of minutes to decades. Some sources are transient, meaning that they undergo intermittent periods of active accretion lasting weeks to months, punctuated by longer quiescent intervals lasting months to years. The cumulative effect of these variations for the current study is that our observational data, which spans 16 years, samples most bursting sources over a wide range of intensities (or accretion rates). 

The MINBAR
observations were performed with one of three satellite-based X-ray
instruments: the Wide Field Cameras (WFC) onboard {\it BeppoSAX}
\cite[]{jager97,zand04b}, the Proportional Counter Array (PCA) onboard
{\it RXTE}\/ \cite[]{xte96,xtecal06}, and the Joint
European X-ray Monitor \cite[JEM-X;][]{lund03} onboard {\it INTEGRAL}. Each instrument covers the energy range 3--25~keV, which was chosen as the common band to measure the X-ray flux $F_X$ for each observation.
The luminosity $L_X$ can be estimated from the X-ray flux $F_X$ (measured in some instrumental band) as 
\begin{equation}
L_X=4\pi d^2 \xi_p F_X c_{\rm bol}
\end{equation}
where $d$ is the distance to the source, $\xi_p$ a parameter that accounts for any anisotropy of the X-ray emission with respect to the system inclination \cite[relative to the line of sight; e.g. ][]{he16}, and $c_{\rm bol}$ the bolometric correction that accounts for the limited passband of the instruments. 

Conventionally, the accretion rate can then be estimated from the inferred X-ray luminosity. For an accretion rate $\dot{M}$ onto a neutron star with mass $M$ and radius $R$, the luminosity (assuming conservative accretion and perfect radiative efficiency) is 
\begin{equation}
L_X=\frac{GM\dot{M}}{R}
\end{equation}
where $G$ is the gravitational constant. 

The distance $d$ to each source may be estimated from the peak flux of 
photospheric radius-expansion (PRE) bursts.
As part of the MINBAR data analysis, we have identified PRE bursts from all sources and measured a mean peak flux $\left<F_{\rm Edd}\right>$ for each source\footnote{These burst peak fluxes are based on time resolved spectroscopy following \cite{bcatalog}, but adopting the revised PCA effective area incorporated into {\sc pcarsp} version 11.7 and correcting for the effects of instrumental deadtime}. 
These values are listed in Table \ref{tab:sources}; note that the quoted uncertainty includes the intrinsic variation of peak intensities of radius-expansion bursts from each source \cite[cf. with][]{bcatalog}.

Rather than calculate the distance, accretion luminosity and hence $\dot{M}$, we can estimate the accretion rate as a fraction of the Eddington value $\dot{M}_{\rm Edd}$ more directly \cite[following][]{vppl88}, by calculating the ratio of bolometric persistent flux to the inferred Eddington flux measured from the photospheric radius-expansion bursts, i.e. 
\begin{equation}
\dot{M}/\dot{M}_{\rm Edd} = \frac{\xi_p F_X c_{\rm bol}}{\xi_b
\left<F_{\rm Edd}\right>}(1+X)
\end{equation}
where $\xi_b$ is the equivalent anisotropy term specific to the burst emission (which is expected to be different from $\xi_p$).
This approach is equivalent to calculating the distance and luminosity, but omits the intermediate steps. Here we include one additional correction factor (constant for all the sources) related to the hydrogen fraction $X$ of the material in the atmosphere. Observational evidence suggests that even for sources that accrete material with solar hydrogen abundance ($X = 0.7$ mass fraction) the corresponding Eddington limit is for material that is deficient in hydrogen. This situation may arise following the ejection of the hydrogen-rich material during the radius-expansion episode \cite[]{gal06a}.

The value of the anisotropy factors $\xi_p$ and $\xi_b$ depend on the
geometry of the accretion disk and the inclination angle $i$ between the
disk and the line-of-sight. For most sources, this angle is not well
known. The exception is for systems where the inclination is sufficiently
high that partial (or complete) X-ray eclipses are observed, arising from
obscuration of the neutron star by material in the disk or the companion.
The only such example in our sample is EXO~0748$-$676, for which the
inclination has been estimated at $i = 75^\circ$ \cite[]{parmar86}. For this system, we adopt a
correction factor to the inferred accretion rate based on recent models of
the effect of the accretion disk \cite[]{he16}, of $\xi_p/\xi_b
=3.21/1.40=2.30$. For the remaining systems, we adopt a correction factor
corresponding to the median expected inclination in the range $i<72^\circ$ \cite[since dipping is not observed consistently; e.g.][]{gal16a} assuming an isotropic distribution of system orientations, $\xi_p/\xi_b =0.809/0.898=0.900$. Because the true inclination for individual non-dipping sources may be anywhere from zero up to this maximum value, the range of error introduced for the accretion rate is a factor of 0.67--2.1.
 
For each observation we measured the flux $F_X$ by performing a model fit to the X-ray spectrum accumulated over the observation (lasting between 30~min and a few days), covering the (common) instrumental energy range 3--25~keV. The specific model for each observation is chosen from a range of phenomenological models depending upon the signal-to-noise and the source state, and range from a simple power-law, power law plus blackbody (Planck spectrum), or a spectrum simulating Comptonisation in the neutron star atmosphere. For the WFC and JEM-X, which have peak effective areas of approximately $100\ {\rm cm}^2$, the relatively low signal-to-noise means that most observations could be well-fit (with reduced $\chi^2 \approx 1$) with a simple power law. For the PCA observations, the much higher effective area ($\approx 6500\ {\rm cm}^2$ with all five proportional counter units operating) meant that a wider range of spectral models were required. The flux measurement depended only weakly on the specific model chosen. 
We also simulated the effects of absorption by neutral material along the line of sight, adopting a column density appropriate for each source based on prior measurements and/or survey observations.

We then applied a bolometric correction $c_{\rm bol}$ to the 3--25~keV flux to account for X-ray emission outside the instrumental energy band. This contribution can be estimated only for the PCA measurements, by also fitting the spectra from the High-Energy X-ray Timing Experiment (HEXTE) instrument, which covers the energy range 16--250 keV (although the target sources are typically detected up to 100 keV at most). As we cannot derive a bolometric correction for every observation, we adopt instead a correction specific to each source, calculated as the average for {\it RXTE} observations within the flux bin in which the maximum burst rate was reached. The overall range of this correction for the sources analysed here is 1.47--2.10.

One caveat affecting the accretion rate as estimated here comes from observations of some transient systems, which suggest that the persistent flux may not always be an unambiguous measure of the true rate. There are several examples that show markedly different burst behavior at similar inferred accretion rates \cite[e.g.][]{chenevez11b}. This apparent ``hysteresis'' in the burst behavior has not yet been demonstrated for persistent sources (which make up 5 of the 8 sources analysed here). Further analysis of the MINBAR sample may serve to test for this behavior.

We performed a separate analysis for the recently-discovered globular cluster source, Terzan~5~X-2. For this source we adopted a lower limit on the accretion rate at which the burst rate reached a maximum, corresponding to the maximum luminosity at which the source was observed during its 2010 outburst \cite[]{linares12a}. Taking into account the uniform isotropy corrections as for the other systems, this value corresponds to $0.45\ \dot{M}_{\rm Edd}$. This object has singular properties within our sample, as it exhibits pulsations at 11~Hz, indicating a spin frequency more than an order of magnitude lower than the next slowest-spinning example. We note that the burst behavior was similar to the next-slowest spinning sources, exhibiting a consistently increasing burst rate with no evidence for a decrease at high accretion rates. However, as the bursts became more frequent, they also became weaker, transitioning to a quasi-periodic variation at mHz timescales, a behavior not seen at comparable accretion rates in other systems. 
This phenomenon is  identified with the onset of quasi-steady He burning, although it occurs in other systems at accretion rates well below the level where burning is expected to become stable \cite[]{rev01}. 

\section{Results} 
\label{sec:results}

\begin{figure}[ht]
\includegraphics[width=0.48\textwidth]{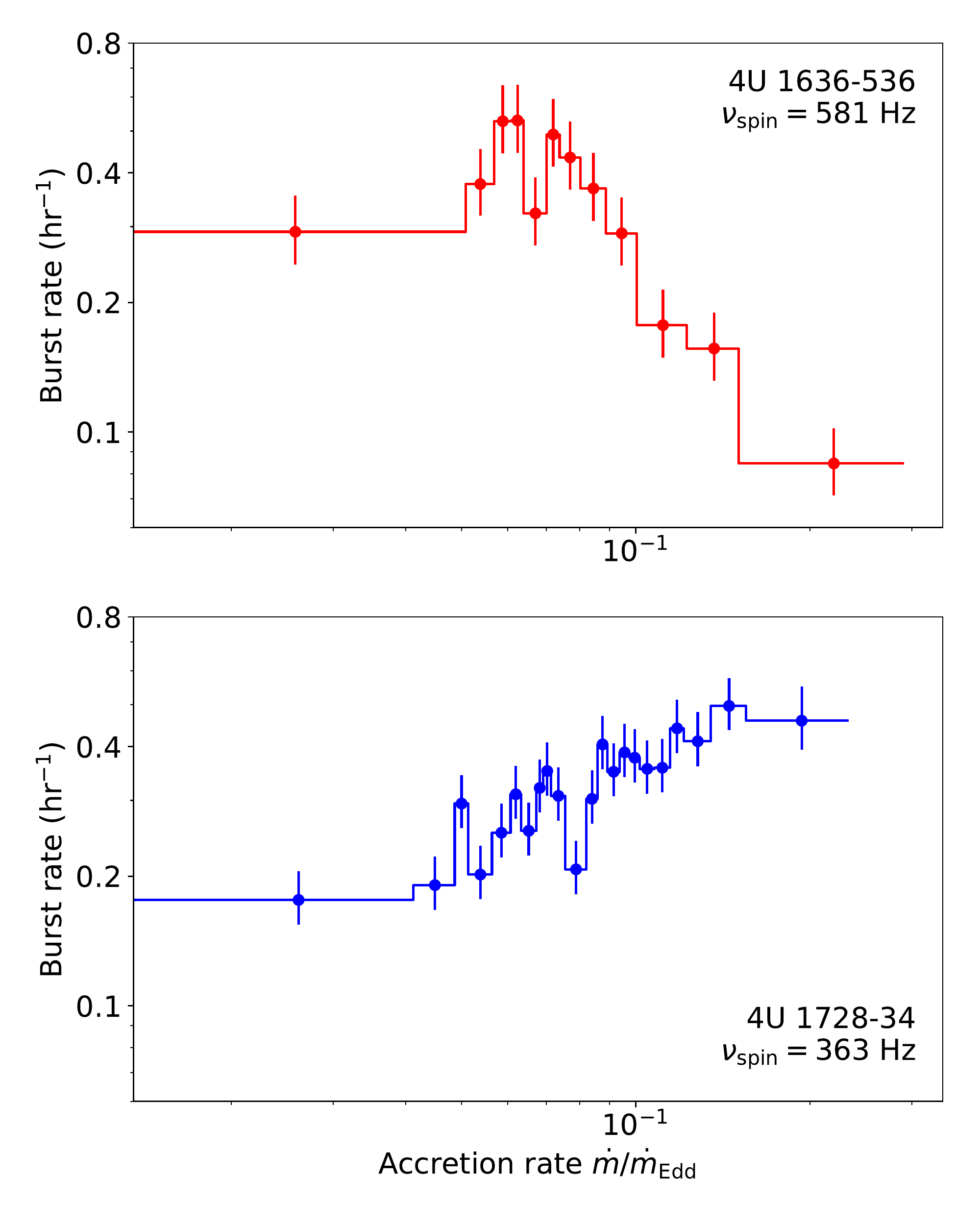} % 4U1636-536_rate_bc2.95.png
\caption{The rate of thermonuclear bursts as a function of inferred accretion rate for selected neutron-star binaries from the MINBAR sample. The top panel shows the measured rate for 
4U~1636$-$536, which reaches a maximum of $0.6\ {\rm hr}^{-1}$ at an accretion rate between 0.04--0.06 of the Eddington rate, while for 4U~1728-34 ({\it bottom panel}), the burst rate increases gradually over the entire range of accretion rates observed. The rate calculated within each accretion rate bin is from 50 bursts, and the uncertainty is estimated assuming Poisson statistics on the number of detected bursts. 
\label{fig:examples}}
\end{figure}

We show examples of the burst rate measured as a function of accretion rate for sources in the MINBAR sample in Figure \ref{fig:examples}; these measurements are broadly consistent with earlier results from smaller sub-samples of the MINBAR data \cite[]{corn03a,bcatalog}. For some sources, the burst rate increases with accretion rate only up to a point; at higher accretion rates the burst rate instead decreases, with the bursts also becoming irregular, and weaker. An example is the persistently-active source 4U 1636$-$536, as shown in Figure \ref{fig:examples} (top panel).
We found no correlation between the maximum burst rate for different sources and the spin period, finding values in the range of (typically) 0.3--$0.6\ {\rm hr}^{-1}$ (excluding Terzan 5 X-2).

Not all the sources studied show the same behavior as 4U~1636$-$536, despite being observed over comparable ranges of accretion rate. For several sources, the burst rate increased steadily up to the highest accretion rate at which the source was observed, and no decrease in burst rate was measured. 

The most compelling example of this alternative behavior is 4U~1728$-$34, the most prolific source overall, with %1094 
more than a thousand 
bursts identified in the MINBAR sample. For this system, measurements in the range 0.02--0.30 times the Eddington accretion rate find the source bursting at a gradually increasing rate, from 0.2--0.5~hr$^{-1}$ (Figure \ref{fig:examples}, bottom panel). While some bin-to-bin variations are present, the behavior is markedly different from 4U~1636$-$536, despite spanning a similar range of accretion rates. 

These two sources differ in several respects. The mass donor in 4U~1636$-$536 is likely hydrogen-rich, 
based on the occasional presence of long-duration bursts indicative of rp-process H-burning, as well as short recurrence time bursts \cite[e.g.][]{bcatalog,keek10}. In
4U~1728$-$34, the donor is probably hydrogen-poor \cite[]{gal06a,gal10b}. Even so, for both types, theoretical models predict a monotonically increasing burst rate with accretion rate \cite[e.g.][]{ramesh03}; there is no known mechanism that would associate the presence of hydrogen in the burst fuel with a falling burst rate. A more interesting difference is the variation of neutron star spin between the two systems, with 4U~1636$-$536 spinning at 581~Hz, substantially above that for 4U~1728$-$34, at 363~Hz \cite[e.g.][]{watts12a}.

\begin{figure}[ht!]
\includegraphics[width=0.48\textwidth]{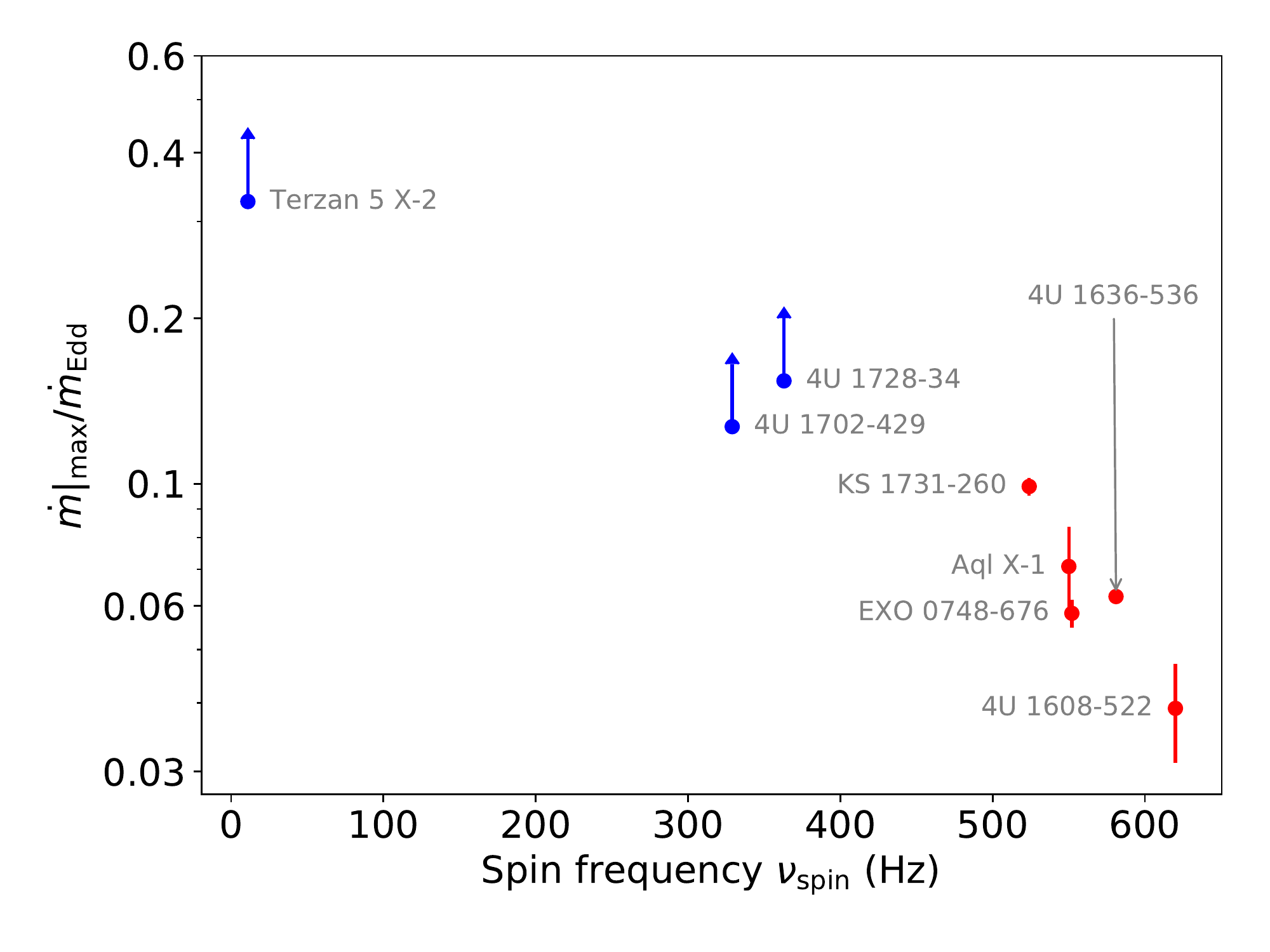}
\caption{The accretion rate $\dot{m}|_{\rm max}$   at which the burst rate
reaches a maximum, plotted as a function of the measured spin of the
bursting neutron star. Note the pronounced anti-correlation with the spin
frequency. For sources with spin frequencies below 500~Hz, we found no
decrease in the burst rate at the highest accretion rates; for these
sources, the $\dot{m}|_{\rm max}$  is a lower limit ({\it blue symbols}). 
The error bars correspond to the width of the accretion-rate bin over which the maximum is measured, and neglect the additional uncertainty arising from the poorly-constrained anisotropy correction factor (for sources other than EXO~0748$-$676) 
and the hydrogen mass fraction $X$ in the accreted fuel.
\label{fig:gammamax}}
\end{figure}

Motivated by the different burst rate behavior for the two most prolific sources, we quantified the burst behavior of 
our target 
sources by measuring the accretion rate $\dot{m}|_{\rm max}$  at which the burst rate achieved a maximum (Figure \ref{fig:gammamax}), including measurements from the literature of the slowest-spinning burst system known, Terzan 5 X-2, as described in \S\ref{sec:obs}.
Our results reveal an anti-correlation of $\dot{m}|_{\rm max}$  on the spin rate. For the slowest-spinning sources, all with spins lower than 400 Hz, we found no evidence of a maximum burst rate. For these systems we adopt the maximum accretion rate at which they were observed as a lower limit to $\dot{m}|_{\rm max}$. In contrast, each of the four sources spinning faster than 500 Hz exhibited a maximum in the burst rate, at an accretion rate that decreased as the spin rate increased.

The observed anti-correlation between $\dot{m}|_{\rm max}$  and the neutron star spin is highly significant, with a Spearman's rank correlation coefficient of $\rho = -0.952$, corresponding with a null hypothesis probability of 0.00026 ($3.5\sigma$). As the three measurements at the lowest spin values are only limits,
it is possible that the true values of $\dot{m}|_{\rm max}$ are {\it positively} correlated with spin rate, against the anti-correlation of the measurements at higher spin rates.
In that case,
we may estimate a minimum significance based on the lowest possible rank correlation of -0.881, of $p = 0.00385$, or $2.7\sigma$. We also considered the possible effects of the unknown system inclination for the sources other than EXO~0748$-$676. By drawing random samples for an isotropic distribution of inclinations, we find that 39\% of the correlation coefficients are less than our conservative limit, with a 95\% upper limit of $\rho = -0.690$, corresponding to $p = 0.058$. Thus, we consider it 
likely that a significant anti-correlation exists between these two quantities.

\section{Discussion} 
\label{sec:discussion}

Numerical models of burst behavior predict that the burst rate increases monotonically with increasing accretion rate, to the point where the burning is stabilized and bursting ceases \cite[e.g.][]{lampe16}.
That the observed burst rate sometimes decreases with increasing accretion rate has been known for some time \cite[]{corn03a}, but the data presented here are the first to suggest a connection with the neutron star spin. 

It is sometimes suggested that decreasing burst rate measurements can be attributed to systematic errors in the determination of the accretion rate \cite[e.g.][]{bil98a}. It is possible that changes in the spectral shape and/or accretion geometry or efficiency may alter this proportionality over the range of accretion rates observed, distorting the position or detailed shape of the burst rate curve.
Empirical estimates of the scale of this uncertainty are approximately 40\% \cite[]{thompson08}.
Another possibility is that, while the {\it global} accretion rate increases, the area over which accretion occurs increases more rapidly, so that the accretion rate per unit area (which sets the burst ignition conditions) also {\it decreases} \cite[]{bil00}.
However, it is unlikely that such effects could result in an inversion of the theoretically predicted curve without major modifications from current models.

A decreasing burst rate may arise from an increasing role for steady burning simultaneously with unsteady burning, perhaps over different regions on the neutron star surface. 
One way to achieve localized stable burning may be via local heating of certain regions on the star, 
while burning remains unstable elsewhere. Numerical simulations suggest that a high degree of heating (equivalent to several MeV/nucleon) is required to decrease the critical accretion rate for stable burning \cite[e.g.][]{keek14b}. If this hypothesis is correct, the results presented here imply that the degree of heating is related to the neutron star rotation. That constraint excludes several possibilities.

First, heating from the shear between the faster-moving accretion disk and the star might be expected to result in heating down to the fuel layer \cite[e.g.][]{is10}. However, the material in the accretion disk is expected to be moving in approximately Keplerian orbits prior to falling on the neutron star, so that the orbital angular velocity will be consistently in excess of the neutron star rotation. Thus, such heating would be greater for slower-rotating stars, the opposite effect that we observe. 

Second, there is observational evidence for additional heat sources operating below the thermonuclear burning layer, as inferred from measurements of cooling of neutron star crusts following accretion episodes \cite[e.g.][]{bc09} 
as well as the ``superburst'' ignition conditions in some sources \cite[]{brown04,cumming06}.
Typical heating values are 1--2~MeV/nucleon, but there is no evidence of a correlation with neutron star spin. 

Alternatively, a direct effect of rotation on the burst ignition arises from the effective gravity, which is up to 25\% smaller at the equator of the star \cite[e.g.][]{slu02}. Burst ignition is expected to occur there preferentially, only moving to higher (or lower) latitudes when the accretion rate approaches the limit at which burning stabilizes \cite[]{mw08}. The dependence of burst rate on accretion rate will change if ignition moves off-equator, with the size of the effect depending on the spin rate. The predicted evolution of the burst rate for current (albeit simple) models of the ignition layer cannot reproduce the behaviour we observe, without additional physics playing a role \cite[]{cavecchi17}.

A possible explanation instead comes in the form of additional diffusivity arising from Eddington-Sweet circulation. In simulations of pure He bursts, this process has been shown to contribute additional mixing of the burst ashes into fresh fuel, thus lowering the critical accretion rate for stability as the rotation rate increases \cite[]{keek09}. Most of the sources in our sample are known to accrete mixed hydrogen and helium fuel, and the predicted influence of rotation on such mixtures are unknown. Simulations of runaways on the surface of white dwarfs have also shown that rotation can indeed suppress ignition, but the angular frequencies considered are $10^{-3}$ of those in neutron stars 
\cite[]{yoon04}. The result presented here further motivates the need for more realistic simulations of the bursting layer, including the effects of rotation, as well as more comprehensive observational studies which can more clearly identify the factors giving rise to variations in burst properties between different sources.

\acknowledgments

We thank the anonymous reviewer whose comments significantly improved this
paper.
The MINBAR project acknowledges the support of the Australian Academy of
Science's Scientific Visits to Europe program, and the Australian Research
Council's Discovery Projects (project DP0880369) and Future Fellowship
(project FT0991598) schemes. The research leading to these results has
received funding from the European Union's Horizon 2020 Programme under
AHEAD project (grant agreement n. 654215). HW acknowledges support by the
German DLR under contracts 50 OR 1405 and 50 OR 1711. LO acknowledges
support from an NWO Top Grant, Module 1 (PI Rudy Wijnands). ALW
acknowledges support from ERC Starting Grant No. 639217 CSINEUTRONSTAR.

\vspace{5mm}
\facilities{BeppoSAX(WFC), RXTE(PCA and HEXTE), INTEGRAL(JEM-X)}

\software{% astropy \citep{astropy13},  
          XSpec \citep{xspec12},
          matplotlib \citep{matplotlib07}
          }

% \bibliography{all}

\begin{thebibliography}{47}
\expandafter\ifx\csname natexlab\endcsname\relax\def\natexlab#1{#1}\fi

\bibitem[{{Bildsten}(1998)}]{bil98a}
{Bildsten}, L. 1998, in The Many Faces of Neutron Stars, ed. R.~{Buccheri},
  J.~{van Paradijs}, \& A.~{Alpar} (Dordrecht: Kluwer), 419

\bibitem[{{Bildsten}(2000)}]{bil00}
{Bildsten}, L. 2000, in Cosmic Explosions, the 10th Annual October Astrophysics
  Conference, Maryland, October 11--13 1999, AIP Conf. 522, ed. S.~{Holt} \&
  W.~{Zhang} (Woodbury NY: AIP), 359--369

\bibitem[{{Brown}(2004)}]{brown04}
{Brown}, E.~F. 2004, \apjl, 614, L57

\bibitem[{{Brown} \& {Cumming}(2009)}]{bc09}
{Brown}, E.~F. \& {Cumming}, A. 2009, \apj, 698, 1020

\bibitem[{{Burke} {et~al.}(2018){Burke}, {Gilfanov}, \& {Sunyaev}}]{burke18}
{Burke}, M.~J., {Gilfanov}, M., \& {Sunyaev}, R. 2018, \mnras, 474, 760

\bibitem[{{Cavecchi} {et~al.}(2017){Cavecchi}, {Watts}, \&
  {Galloway}}]{cavecchi17}
{Cavecchi}, Y., {Watts}, A.~L., \& {Galloway}, D.~K. 2017, \apj, 851, 1

\bibitem[{{Chakrabarty} {et~al.}(2003){Chakrabarty}, {Morgan}, {Muno},
  {Galloway}, {Wijnands}, {van der Klis}, \& {Markwardt}}]{chak03a}
{Chakrabarty}, D., {Morgan}, E.~H., {Muno}, M.~P., {Galloway}, D.~K.,
  {Wijnands}, R., {van der Klis}, M., \& {Markwardt}, C.~B. 2003, \nat, 424, 42

\bibitem[{{Chenevez} {et~al.}(2011){Chenevez}, {Altamirano}, {Galloway}, {in't
  Zand}, {Kuulkers}, {Degenaar}, {Falanga}, {Del Monte}, {Evangelista},
  {Feroci}, \& {Costa}}]{chenevez11b}
{Chenevez}, J., {Altamirano}, D., {Galloway}, D.~K., {in't Zand}, J.~J.~M.,
  {Kuulkers}, E., {Degenaar}, N., {Falanga}, M., {Del Monte}, E.,
  {Evangelista}, Y., {Feroci}, M., \& {Costa}, E. 2011, \mnras, 410, 179

\bibitem[{{Clark} {et~al.}(1976){Clark}, {Jernigan}, {Bradt}, {Canizares},
  {Lewin}, {Li}, {Mayer}, {McClintock}, \& {Schnopper}}]{clark76}
{Clark}, G.~W., {Jernigan}, J.~G., {Bradt}, H., {Canizares}, C., {Lewin},
  W.~H.~G., {Li}, F.~K., {Mayer}, W., {McClintock}, J., \& {Schnopper}, H.
  1976, \apjl, 207, L105

\bibitem[{{Cornelisse} {et~al.}(2003){Cornelisse}, {in 't Zand}, {Verbunt},
  {Kuulkers}, {Heise}, {den Hartog}, {Cocchi}, {Natalucci}, {Bazzano}, \&
  {Ubertini}}]{corn03a}
{Cornelisse}, R., {in 't Zand}, J.~J.~M., {Verbunt}, F., {Kuulkers}, E.,
  {Heise}, J., {den Hartog}, P.~R., {Cocchi}, M., {Natalucci}, L., {Bazzano},
  A., \& {Ubertini}, P. 2003, \aap, 405, 1033

\bibitem[{{Cumming} {et~al.}(2006){Cumming}, {Macbeth}, {Zand}, \&
  {Page}}]{cumming06}
{Cumming}, A., {Macbeth}, J., {Zand}, J.~J.~M.~i., \& {Page}, D. 2006, \apj,
  646, 429

\bibitem[{{Darnley} {et~al.}(2016){Darnley}, {Henze}, {Bode}, {Hachisu},
  {Hernanz}, {Hornoch}, {Hounsell}, {Kato}, {Ness}, {Osborne}, {Page},
  {Ribeiro}, {Rodr{\'{\i}}guez-Gil}, {Shafter}, {Shara}, {Steele}, {Williams},
  {Arai}, {Arcavi}, {Barsukova}, {Boumis}, {Chen}, {Fabrika}, {Figueira},
  {Gao}, {Gehrels}, {Godon}, {Goranskij}, {Harman}, {Hartmann}, {Hosseinzadeh},
  {Horst}, {Itagaki}, {Jos{\'e}}, {Kabashima}, {Kaur}, {Kawai}, {Kennea},
  {Kiyota}, {Ku{\v c}{\'a}kov{\'a}}, {Lau}, {Maehara}, {Naito}, {Nakajima},
  {Nishiyama}, {O'Brien}, {Quimby}, {Sala}, {Sano}, {Sion}, {Valeev},
  {Watanabe}, {Watanabe}, {Williams}, \& {Xu}}]{darnley16}
{Darnley}, M.~J., {Henze}, M., {Bode}, M.~F., {Hachisu}, I., {Hernanz}, M.,
  {Hornoch}, K., {Hounsell}, R., {Kato}, M., {Ness}, J.-U., {Osborne}, J.~P.,
  {Page}, K.~L., {Ribeiro}, V.~A.~R.~M., {Rodr{\'{\i}}guez-Gil}, P., {Shafter},
  A.~W., {Shara}, M.~M., {Steele}, I.~A., {Williams}, S.~C., {Arai}, A.,
  {Arcavi}, I., {Barsukova}, E.~A., {Boumis}, P., {Chen}, T., {Fabrika}, S.,
  {Figueira}, J., {Gao}, X., {Gehrels}, N., {Godon}, P., {Goranskij}, V.~P.,
  {Harman}, D.~J., {Hartmann}, D.~H., {Hosseinzadeh}, G., {Horst}, J.~C.,
  {Itagaki}, K., {Jos{\'e}}, J., {Kabashima}, F., {Kaur}, A., {Kawai}, N.,
  {Kennea}, J.~A., {Kiyota}, S., {Ku{\v c}{\'a}kov{\'a}}, H., {Lau}, K.~M.,
  {Maehara}, H., {Naito}, H., {Nakajima}, K., {Nishiyama}, K., {O'Brien},
  T.~J., {Quimby}, R., {Sala}, G., {Sano}, Y., {Sion}, E.~M., {Valeev}, A.~F.,
  {Watanabe}, F., {Watanabe}, M., {Williams}, B.~F., \& {Xu}, Z. 2016, \apj,
  833, 149

\bibitem[{{Done} {et~al.}(2007){Done}, {Gierli{\'n}ski}, \& {Kubota}}]{done07}
{Done}, C., {Gierli{\'n}ski}, M., \& {Kubota}, A. 2007, \aapr, 15, 1

\bibitem[{{Dorman} \& {Arnaud}(2001)}]{xspec12}
{Dorman}, B. \& {Arnaud}, K.~A. 2001, in Astronomical Society of the Pacific
  Conference Series, Vol. 238, Astronomical Data Analysis Software and Systems
  X, ed. {F.~R.~Harnden Jr., F.~A.~Primini, \& H.~E.~Payne}, 415

\bibitem[{{Fujimoto}(1993)}]{fuji93}
{Fujimoto}, M.~Y. 1993, \apj, 419, 768

\bibitem[{{Galloway} {et~al.}(2016){Galloway}, {Ajamyan}, {Upjohn}, \&
  {Stuart}}]{gal16a}
{Galloway}, D.~K., {Ajamyan}, A.~N., {Upjohn}, J., \& {Stuart}, M. 2016,
  \mnras, 461, 3847

\bibitem[{{Galloway} \& {Keek}(2017)}]{gal17b}
{Galloway}, D.~K. \& {Keek}, L. 2017, ArXiv e-prints

\bibitem[{{Galloway} {et~al.}(2008){Galloway}, {Muno}, {Hartman}, {Psaltis}, \&
  {Chakrabarty}}]{bcatalog}
{Galloway}, D.~K., {Muno}, M.~P., {Hartman}, J.~M., {Psaltis}, D., \&
  {Chakrabarty}, D. 2008, \apjs, 179, 360

\bibitem[{{Galloway} {et~al.}(2006){Galloway}, {Psaltis}, {Muno}, \&
  {Chakrabarty}}]{gal06a}
{Galloway}, D.~K., {Psaltis}, D., {Muno}, M.~P., \& {Chakrabarty}, D. 2006,
  \apj, 639, 1033

\bibitem[{{Galloway} {et~al.}(2010){Galloway}, {Yao}, {Marshall}, {Misanovic},
  \& {Weinberg}}]{gal10b}
{Galloway}, D.~K., {Yao}, Y., {Marshall}, H., {Misanovic}, Z., \& {Weinberg},
  N. 2010, \apj, 724, 417

\bibitem[{{Grindlay} {et~al.}(1976){Grindlay}, {Gursky}, {Schnopper},
  {Parsignault}, {Heise}, {Brinkman}, \& {Schrijver}}]{grindlay76}
{Grindlay}, J., {Gursky}, H., {Schnopper}, H., {Parsignault}, D.~R., {Heise},
  J., {Brinkman}, A.~C., \& {Schrijver}, J. 1976, \apjl, 205, L127

\bibitem[{{He} \& {Keek}(2016)}]{he16}
{He}, C.-C. \& {Keek}, L. 2016, \apj, 819, 47

\bibitem[{Hunter(2007)}]{matplotlib07}
Hunter, J.~D. 2007, Computing In Science \& Engineering, 9, 90

\bibitem[{{in 't Zand} {et~al.}(2004){in 't Zand}, {Verbunt}, {Heise},
  {Bazzano}, {Cocchi}, {Cornelisse}, {Kuulkers}, {Natalucci}, \&
  {Ubertini}}]{zand04b}
{in 't Zand}, J.~J.~M., {Verbunt}, F., {Heise}, J., {Bazzano}, A., {Cocchi},
  M., {Cornelisse}, R., {Kuulkers}, E., {Natalucci}, L., \& {Ubertini}, P.
  2004, in Proceedings of the 2nd BeppoSAX Conference: "The Restless
  High-Energy Universe", Amsterdam, 5--9 May 2003, ed. E.~P.~J. {van den
  Heuvel}, R.~A.~M.~J. {Wijers}, \& J.~J.~M. {in 't Zand}, Vol. 132, 486--495

\bibitem[{{Inogamov} \& {Sunyaev}(2010)}]{is10}
{Inogamov}, N.~A. \& {Sunyaev}, R.~A. 2010, Astronomy Letters, 36, 848

\bibitem[{{Jager} {et~al.}(1997){Jager}, {Mels}, {Brinkman}, {Galama},
  {Goulooze}, {Heise}, {Lowes}, {Muller}, {Naber}, {Rook}, {Schuurhof},
  {Schuurmans}, \& {Wiersma}}]{jager97}
{Jager}, R., {Mels}, W.~A., {Brinkman}, A.~C., {Galama}, M.~Y., {Goulooze}, H.,
  {Heise}, J., {Lowes}, P., {Muller}, J.~M., {Naber}, A., {Rook}, A.,
  {Schuurhof}, R., {Schuurmans}, J.~J., \& {Wiersma}, G. 1997, \aaps, 125, 557

\bibitem[{{Jahoda} {et~al.}(2006){Jahoda}, {Markwardt}, {Radeva}, {Rots},
  {Stark}, {Swank}, {Strohmayer}, \& {Zhang}}]{xtecal06}
{Jahoda}, K., {Markwardt}, C.~B., {Radeva}, Y., {Rots}, A.~H., {Stark}, M.~J.,
  {Swank}, J.~H., {Strohmayer}, T.~E., \& {Zhang}, W. 2006, \apjs, 163, 401

\bibitem[{{Jahoda} {et~al.}(1996){Jahoda}, {Swank}, {Giles}, {Stark},
  {Strohmayer}, {Zhang}, \& {Morgan}}]{xte96}
{Jahoda}, K., {Swank}, J.~H., {Giles}, A.~B., {Stark}, M.~J., {Strohmayer}, T.,
  {Zhang}, W., \& {Morgan}, E.~H. 1996, \procspie, 2808, 59

\bibitem[{{Jos{\'e}} \& {Hernanz}(2007)}]{jose07}
{Jos{\'e}}, J. \& {Hernanz}, M. 2007, Journal of Physics G Nuclear Physics, 34,
  R431

\bibitem[{{Keek} {et~al.}(2014){Keek}, {Cyburt}, \& {Heger}}]{keek14b}
{Keek}, L., {Cyburt}, R.~H., \& {Heger}, A. 2014, \apj, 787, 101

\bibitem[{{Keek} {et~al.}(2010){Keek}, {Galloway}, {in't Zand}, \&
  {Heger}}]{keek10}
{Keek}, L., {Galloway}, D.~K., {in't Zand}, J.~J.~M., \& {Heger}, A. 2010,
  \apj, 718, 292

\bibitem[{{Keek} {et~al.}(2009){Keek}, {Langer}, \& {in't Zand}}]{keek09}
{Keek}, L., {Langer}, N., \& {in't Zand}, J.~J.~M. 2009, \aap, 502, 871

\bibitem[{{Lampe} {et~al.}(2016){Lampe}, {Heger}, \& {Galloway}}]{lampe16}
{Lampe}, N., {Heger}, A., \& {Galloway}, D.~K. 2016, \apj, 819, 46

\bibitem[{{Lewin} {et~al.}(1993){Lewin}, {van Paradijs}, \& {Taam}}]{lew93}
{Lewin}, W. H.~G., {van Paradijs}, J., \& {Taam}, R.~E. 1993, \ssr, 62, 223

\bibitem[{{Linares} {et~al.}(2012){Linares}, {Altamirano}, {Chakrabarty},
  {Cumming}, \& {Keek}}]{linares12a}
{Linares}, M., {Altamirano}, D., {Chakrabarty}, D., {Cumming}, A., \& {Keek},
  L. 2012, \apj, 748, 82

\bibitem[{{Lund} {et~al.}(2003){Lund}, {Budtz-J{\o}rgensen}, {Westergaard},
  {Brandt}, {Rasmussen}, {Hornstrup}, {Oxborrow}, {Chenevez}, {Jensen},
  {Laursen}, {Andersen}, {Mogensen}, {Rasmussen}, {Om{\o}}, {Pedersen},
  {Polny}, {Andersson}, {Andersson}, {K{\"a}m{\"a}r{\"a}inen}, {Vilhu},
  {Huovelin}, {Maisala}, {Morawski}, {Juchnikowski}, {Costa}, {Feroci},
  {Rubini}, {Rapisarda}, {Morelli}, {Carassiti}, {Frontera}, {Pelliciari},
  {Loffredo}, {Mart{\'{\i}}nez N{\'u}{\~n}ez}, {Reglero}, {Velasco}, {Larsson},
  {Svensson}, {Zdziarski}, {Castro-Tirado}, {Attina}, {Goria}, {Giulianelli},
  {Cordero}, {Rezazad}, {Schmidt}, {Carli}, {Gomez}, {Jensen}, {Sarri},
  {Tiemon}, {Orr}, {Much}, {Kretschmar}, \& {Schnopper}}]{lund03}
{Lund}, N., {Budtz-J{\o}rgensen}, C., {Westergaard}, N.~J., {Brandt}, S.,
  {Rasmussen}, I.~L., {Hornstrup}, A., {Oxborrow}, C.~A., {Chenevez}, J.,
  {Jensen}, P.~A., {Laursen}, S., {Andersen}, K.~H., {Mogensen}, P.~B.,
  {Rasmussen}, I., {Om{\o}}, K., {Pedersen}, S.~M., {Polny}, J., {Andersson},
  H., {Andersson}, T., {K{\"a}m{\"a}r{\"a}inen}, V., {Vilhu}, O., {Huovelin},
  J., {Maisala}, S., {Morawski}, M., {Juchnikowski}, G., {Costa}, E., {Feroci},
  M., {Rubini}, A., {Rapisarda}, M., {Morelli}, E., {Carassiti}, V.,
  {Frontera}, F., {Pelliciari}, C., {Loffredo}, G., {Mart{\'{\i}}nez
  N{\'u}{\~n}ez}, S., {Reglero}, V., {Velasco}, T., {Larsson}, S., {Svensson},
  R., {Zdziarski}, A.~A., {Castro-Tirado}, A., {Attina}, P., {Goria}, M.,
  {Giulianelli}, G., {Cordero}, F., {Rezazad}, M., {Schmidt}, M., {Carli}, R.,
  {Gomez}, C., {Jensen}, P.~L., {Sarri}, G., {Tiemon}, A., {Orr}, A., {Much},
  R., {Kretschmar}, P., \& {Schnopper}, H.~W. 2003, \aap, 411, L231

\bibitem[{{Maurer} \& {Watts}(2008)}]{mw08}
{Maurer}, I. \& {Watts}, A.~L. 2008, \mnras, 383, 387

\bibitem[{{Muno} {et~al.}(2004){Muno}, {Galloway}, \& {Chakrabarty}}]{muno04a}
{Muno}, M.~P., {Galloway}, D.~K., \& {Chakrabarty}, D. 2004, \apj, 608, 930

\bibitem[{{Narayan} \& {Heyl}(2003)}]{ramesh03}
{Narayan}, R. \& {Heyl}, J.~S. 2003, \apj, 599, 419

\bibitem[{{Parmar} {et~al.}(1986){Parmar}, {White}, {Giommi}, \&
  {Gottwald}}]{parmar86}
{Parmar}, A.~N., {White}, N.~E., {Giommi}, P., \& {Gottwald}, M. 1986, \apj,
  308, 199

\bibitem[{{Piro} \& {Bildsten}(2007)}]{pb07}
{Piro}, A.~L. \& {Bildsten}, L. 2007, \apj, 663, 1252

\bibitem[{{Revnivtsev} {et~al.}(2001){Revnivtsev}, {Churazov}, {Gilfanov}, \&
  {Sunyaev}}]{rev01}
{Revnivtsev}, M., {Churazov}, E., {Gilfanov}, M., \& {Sunyaev}, R. 2001, \aap,
  372, 138

\bibitem[{{Spitkovsky} {et~al.}(2002){Spitkovsky}, {Levin}, \&
  {Ushomirsky}}]{slu02}
{Spitkovsky}, A., {Levin}, Y., \& {Ushomirsky}, G. 2002, \apj, 566, 1018

\bibitem[{{Thompson} {et~al.}(2008){Thompson}, {Galloway}, {Rothschild}, \&
  {Homer}}]{thompson08}
{Thompson}, T.~W.~J., {Galloway}, D.~K., {Rothschild}, R.~E., \& {Homer}, L.
  2008, \apj, 681, 506

\bibitem[{{van Paradijs} {et~al.}(1988){van Paradijs}, {Penninx}, \&
  {Lewin}}]{vppl88}
{van Paradijs}, J., {Penninx}, W., \& {Lewin}, W.~H.~G. 1988, \mnras, 233, 437

\bibitem[{{Watts}(2012)}]{watts12a}
{Watts}, A.~L. 2012, \araa, 50, 609

\bibitem[{{Yoon} {et~al.}(2004){Yoon}, {Langer}, \& {Scheithauer}}]{yoon04}
{Yoon}, S.-C., {Langer}, N., \& {Scheithauer}, S. 2004, \aap, 425, 217

\end{thebibliography}
% \bibliographystyle{apj}

\end{document}